\documentclass[mathleft]{an}
\usepackage{graphicx,times}
\usepackage{bm}
\newcommand{\EQ}{\begin{equation}}
\newcommand{\EN}{\end{equation}}
\newcommand{\EQA}{\begin{eqnarray}}
\newcommand{\ENA}{\end{eqnarray}}

\newcommand{\Fig}[1]{Fig.~\ref{#1}}

\newcommand{\Figs}[2]{Figs.~\ref{#1} and \ref{#2}}

\newcommand{\bra}[1]{\langle #1\rangle}

{}
{}
{}
\newcommand{\meanemf}{\overline{\cal E} {}}

{}
{}
\newcommand{\meanEMF}{\overline{\mbox{\boldmath ${\cal E}$}}{}}{}
{}
{}
{}
{}
{}
\newcommand{\meanBB}{\overline{\mbox{\boldmath $B$}}{}}{}
{}
{}
{}
{}
{}
{}
{}
{}

{}
{}
{}

\newcommand{\meanB}{\overline{B}}

\newcommand{\meanU}{\overline{U}}

\newcommand{\meanJ}{\overline{J}}

{}

{}
{}

\newcommand{\xxx}{\hat{\mbox{\boldmath $x$}} {}}

\newcommand{\rr}{\mbox{\boldmath $r$} {}}

\newcommand{\uu}{\mbox{\boldmath $u$} {}}
\newcommand{\UU}{\mbox{\boldmath $U$} {}}

\newcommand{\bb}{\mbox{\boldmath $b$} {}}
\newcommand{\BB}{\mbox{\boldmath $B$} {}}

\newcommand{\JJ}{\mbox{\boldmath $J$} {}}

\newcommand{\AAA}{\mbox{\boldmath $A$} {}}

\newcommand{\FF}{\mbox{\boldmath $F$} {}}

\newcommand{\grav}{\mbox{\boldmath $g$} {}}
\newcommand{\nab}{\mbox{\boldmath $\nabla$} {}}
\newcommand{\SSSS}{\mbox{\boldmath ${\sf S}$} {}}

\newcommand{\dd}{{\rm d} {}}
\newcommand{\dV}{\,{\rm d}V {}}
\newcommand{\dS}{\,{\rm d}{{\bm{S}}} {}}

\def\Pra{\mbox{\rm Pr}}

\def\Pm{P_{\rm m}}
\def\Rm{R_{\rm m}}

\def\Lu{\mbox{\rm Lu}}
\def\csz{c_{\rm s0}}
\def\cs{c_{\rm s}}

\def\kf{k_{\rm f}}

\def\vArms{v_{\rm A,rms}}
\def\Brms{B_{\rm rms}}
\def\Urms{U_{\rm rms}}
\def\urms{u_{\rm rms}}

\def\half{{\textstyle{1\over2}}}

\def\onethird{{\textstyle{1\over3}}}

\newcommand{\pd}{\partial}

\newcommand{\DIV}{\vec{\nabla} \cdot }

\newcommand{\meanv}[1]{\overline{\bm #1}}
\Pagespan{797}{}
\Yearpublication{2009}
\Yearsubmission{2009}
\Month{7}
\Volume{330}
\Issue{8}
\DOI{10.1002/asna.200911242}
\title{Shear-driven magnetic buoyancy oscillations}
\author{Violaine Vermersch\inst{1,2}\and Axel
Brandenburg\inst{2,3}\fnmsep\thanks{Corresponding author: brandenb@nordita.org}}
\institute{
Centre de Physique Th\'eorique, \'Ecole Polytechnique,
F-91128 Palaiseau cedex, France
\and
NORDITA, AlbaNova University Center, Roslagstullsbacken 23,
SE 10691 Stockholm, Sweden
\and
Department of Astronomy, AlbaNova University Center,
Stockholm University, SE 10691 Stockholm, Sweden
}
\received{2009 Jul 11} \accepted{2009 Aug 19} \publonline{2009 Sep 20}
\keywords{instabilities -- magnetohydrodynamics (MHD) -- turbulence}
\abstract{%
The effects of uniform horizontal shear on a stably stratified layer of
gas is studied.
The system is initially destabilized by a magnetically buoyant flux tube
pointing in the cross-stream direction.
The shear amplifies the initial field to Lundquist numbers of about
200--400, but then its value drops to about 100--300, depending on
the value of the sub-adiabatic gradient.
The larger values correspond to cases where the stratification is strongly
stable and nearly isothermal.
At the end of the runs the magnetic field is nearly axisymmetric, i.e.\
uniform in the streamwise direction.
In view of Cowling's theorem the sustainment of the field remains a puzzle
and may be due to subtle numerical effects that have not yet been identified
in detail.
In the final state the strength of the magnetic field decreases with
height in such a way that the field is expected to be unstable.
Low amplitude oscillations are seen in the vertical velocity
even at late times, suggesting that they might be persistent.
}

\begin{document}
\maketitle

\section{Introduction}

Dynamos convert kinetic energy into magnetic energy.
A typical example is thermally driven convection.
If the difference between heating and cooling across the domain
is strong enough, thermal energy can be converted into kinetic
energy by the Rayleigh-B\'enard instability, and part of this kinetic
energy can then be converted further into magnetic energy
by the dynamo instability.
However, if there is shear, then this can supply a major part
of the kinetic energy.
In view of Cowling's (1933) anti-dynamo theorem, it is clear that the
magnetic field must be fully three-dimensional.
This is easily achieved if the motions are three-dimensional as well.
A linear shear flow is just one-dimensional, although it is in principle
possible that this flow becomes nonlinearly unstable and develops fully
three-dimensional turbulence.
Unfortunately, this requires rather large Reynolds numbers and is not
easy to achieve.
However, in a stratified system it is also possible to produce
three-dimensional motions by the magnetic buoyancy instability.
This was an important agent responsible for driving a dynamo in the model
of Cline et al.\ (2003).
Yet another possibility is the magneto-rotational instability, which
can also produce three-dimensional motions to drive a dynamo
(Brandenburg et al.\ 1995; Hawley et al.\ 1996; Stone et al.\ 1996),
but this requires that there is also rotation.

In this paper we investigate a system similar to that of
Cline et al.\ (2003), but with a linear shear profile instead
of a sinusoidal one, or rather a modified sinusoidal one where one
flank is steeper than the other.
Another difference is that in our model the shear extends through
all layers and is not localized near the bottom of the domain,
as in the model of Cline et al.\ (2003).
Their setup was motivated by the presence of a strong shear layer
in the solar tachocline (Hughes et al.\ 2007).
However, here we are interested in more general aspects rather than
particular applications.
It is possible that some important dynamics would be lost by ignoring
the additional vertical dependence of shear, but this is not well
understood at present.
Regardless of whether or not self-sustained dynamo action exists,
there are a number of issues that deserve to be addressed in such a setup.
Firstly, it is useful to determine the energy fluxes between kinetic
and magnetic energies and how they are coupled to the shear, for example.
Secondly, if magnetic buoyancy plays a role in producing turbulence,
it should be possible to quantify this by measuring suitable correlations.
In particular, it is not clear whether a stratification that is close to
marginally stable is advantageous compared to one that is strongly stable.
Finally, in order to assess the possibility of large-scale dynamo
action, one needs to determine the turbulent transport coefficients.

\section{The model}

\subsection{Governing equations}

We consider a Cartesian domain with externally imposed linear shear
and vertical gravity leading to density stratification in the $z$ direction.
The full set of hydromagnetic equations for the magnetic vector potential
$\bm{A}$, the density $\rho$, the velocity $\bm{U}$, and the specific
internal energy $e$, can then be written in the form
\begin{equation}
\frac{\mathcal{D} \bm A}{\mathcal{D}t}
= -S A_y \hat{\bm{x}} - (\bm{\nabla}\bm{U})^{\rm T}\bm{A}-\mu_0\eta {\bm J},
\label{equ:AA}
\end{equation}
\begin{equation}
\frac{\mathcal{D} \ln \rho}{\mathcal{D}t} = -\DIV{\bm U},
\end{equation}
\begin{equation}
\rho\frac{\mathcal{D} \bm U}{\mathcal{D}t}
= -S\rho U_x\bm{\hat{y}}-{\bm \nabla}p + \rho{\bm g}
+ \bm{J} \times {\bm B}
+ \bm{\nabla} \cdot 2 \nu \rho \mbox{\boldmath ${\sf S}$},
\label{equ:UU}
\end{equation}
\begin{equation}
\rho\frac{\mathcal{D} e}{\mathcal{D}t} = - p\DIV {\bm U}
+ \bm{\nabla} \cdot K \bm{\nabla}T
+ 2 \rho\nu \mbox{\boldmath ${\sf S}$}^2
+ \mu_0\eta \bm{J}^2,
\label{equ:ene}
 \end{equation}
where $\mathcal{D}/\mathcal{D}t = \pd/\pd t + (\bm{U} + \meanv{U}_0)
\cdot \bm{\nabla}$ is the advective derivative,
$\meanv{U}_0 = (0,Sx,0)$ is the imposed large-scale shear flow,
${\sf S}_{ij}=\half (U_{i,j}+U_{j,i})-\onethird\delta_{ij}\DIV\bm{U}$
is the traceless rate of strain tensor,
$\bm{B} = \bm{\nabla}\! \times\! \bm{A}$ is the magnetic field,
$\bm{J} =\bm{\nabla}\! \times\! \bm{B}/\mu_0$ is the current density,
$\mu_0$ is the magnetic permeability, $\eta$ and $\nu$ are respectively
the magnetic diffusivity and kinematic viscosity, $K$ is the heat conductivity,
and $\bm{g} = -g\hat{\bm{z}}$ is the gravitational acceleration.
The fluid obeys an ideal gas law $p=\rho e (\gamma-1)$, where $p$
is the pressure, and $\gamma = c_{\rm p}/c_{\rm v} = 5/3$ is the ratio
of specific heats at constant pressure and volume, respectively.
The internal energy per unit mass is related to the
temperature via $e=c_{\rm v} T$.

\subsection{Initial and boundary conditions}

Our initial stratification is a polytrope where $p$
is proportional to $\rho^\Gamma$, and
$\Gamma$ is related to the polytropic index $m$ via $\Gamma=1+1/m$.
The superadiabatic gradient is usually defined as the normalized entropy
gradient with respect to logarithmic pressure,
$\nabla-\nabla_{\rm ad}=\dd(s/c_{\rm p})/\dd\ln p$, i.e.\
\EQ
\nabla-\nabla_{\rm ad}=\gamma^{-1}-\Gamma^{-1}.
\EN
A stable stratification corresponds to ${\nabla-\nabla_{\rm ad}<0}$,
i.e.\ $\Gamma<\gamma$.
Following Cline et al.\ (2003) we adopt $m=1.6$, which gives
$\nabla-\nabla_{\rm ad}\approx-0.015$.
(We note that the $\nabla-\nabla_{\rm ad}$ quoted by Cline et al.\ (2003)
is scaled by a factor $2(m+1)$, giving $-0.08$.)

In a polytrope, the temperature, and hence the square of the sound
speed are proportional to the negative gravity potential.
Using $c_{\rm s}^2=(\gamma{-}1)h$ we find
\begin{equation}
  c_{\rm s}^2 = -\frac{\gamma}{m{+}1}\,\Phi,
\end{equation}
where $\Phi=-(z_\infty-z)g$ is the gravitational potential,
and the $z_\infty>z$ is the top of the atmosphere where temperature,
density and pressure would vanish.
In order that $\rho=\rho_0$ and $c_{\rm s}^2=c_{\rm s0}^2$ at
a certain reference height $z=-H_0$, we choose
\begin{equation}\label{zinfty}
  z_\infty = -H_0 + (m{+}1) \frac{c_{\rm s0}^2}{\gamma g}.
\end{equation}
We consider a computational domain with
horizontal extent $-2<x/H_0<2$, $-2<y/H_0<2$, and vertical extent
$-4.12\leq z/H_0\leq -1$, where $z/H_0=-1$ corresponds to the top
of the layer.

Our initial magnetic field is given by a flux tube pointing
in the $x$ direction, but with a perturbation in the $x$ direction.
Thus, our vector potential is given by
\EQ
\AAA=B_0{1+\epsilon\cos k_xx\over1+\delta\rr^2/R^2}\,\xxx\times\delta\rr,
\EN
where $\delta\rr=(x,y,z-z_0)$ is the distance from the core of the
tube in the $yz$ plane.
We choose $B_0=0.1\sqrt{gH_0\rho_0\mu_0}$,
$\epsilon=0.3$, $R=0.2H_0$, and $z_0=-3H_0$.

In all cases we use stress-free boundary conditions for the velocity,
\begin{equation}
U_{x,z} = U_{y,z} = U_z = 0,
\end{equation}
together with a vertical field condition, i.e.\ 
\begin{eqnarray}
B_x = B_y &=& 0.
\end{eqnarray}
For the specific entropy we use either an extrapolating boundary condition,
which allows the values of thermodynamic variables on the two boundaries to
change freely (Run~A), or we fix the energy influx at the bottom and
the temperature at the top (Run~B).
The latter condition is also used by Cline et al.\ (2003).

\subsection{Units and control parameters}

Non-dimensional quantities are obtained by setting
\begin{eqnarray}
d = g = \rho_0 = c_{\rm p} = \mu_0 = 1,
\end{eqnarray}
where $\rho_0$ is the initial density at $z_{\rm top}$.
The units of length, time,
velocity, density, entropy, and magnetic field are
\begin{eqnarray}
[x] & = & H_0\,,\;\; [t] = \sqrt{H_0/g}\,,\;\; [U]=\sqrt{gH_0}\,,\;\;
    [\rho]=\rho_0\,, \nonumber \\   
{[s]} & = & c_{\rm p}\,,\;\; 
     [B]=\sqrt{gH_0\rho_0\mu_0}. 
\end{eqnarray}
We define the fluid and magnetic Prandtl numbers as
\begin{eqnarray}
\Pra=\frac{\nu}{\chi_0},\quad
\Pm=\frac{\nu}{\eta},\quad
\end{eqnarray}
where $\chi_0 = K/(\rho_{\rm m} c_{\rm p})$ is the thermal
diffusivity, and $\rho_0$ is the density at $z=-H_0$.
We define the magnetic Reynolds number and the shear parameter via
\begin{eqnarray}
{\rm Rm} = \frac{\urms}{\eta \kf},\quad
{\rm Sh} = \frac{S}{\urms \kf},
\end{eqnarray}
where $\kf = 2\pi/H_0$ is assumed as a reasonable estimate
for the wavenumber of the energy-carrying eddies.

The simulations were performed with the {\sc Pencil Code}%
\footnote{{http://www.nordita.org/software/pencil-code/}},
which uses sixth-order explicit finite differences in space and third
order accurate time stepping method.
The dependent variables are $\AAA$, $\ln\rho$, $\UU$, and the
specific entropy $s$, which is related to the speed of sound $\cs$ via
\EQ
\cs^2=\csz^2[(\gamma-1)\ln(\rho/\rho_0)+\gamma s/c_{\rm p}],
\EN
where $\csz$ and $\rho_0$ are normalization constants.

\section{General features of the simulation results}

In the following we discuss Runs~A and B that differ only in the boundary
conditions adopted for the entropy.
For Run~A we use extrapolating boundary conditions for the specific entropy,
while for Run~B we fix the energy influx at the bottom and the temperature
at the top.
In all cases we use the same polytropic initial condition.

\begin{figure}[t!]%\begin{center}
\includegraphics[width=\columnwidth]{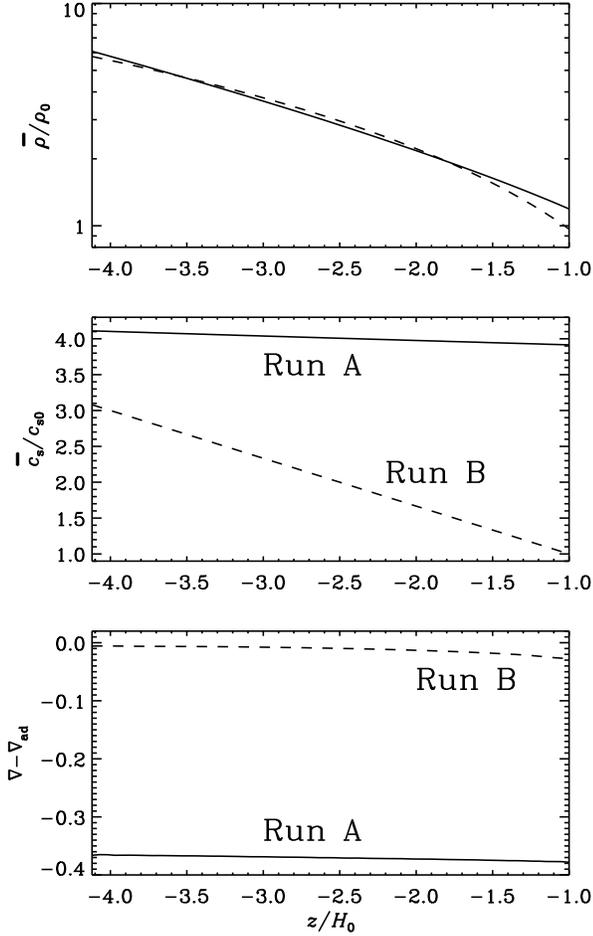}
%\end{center}
\caption[]{
Final stratification for Runs~A (solid lines) and B (dashed lines).
Note that Run~A is nearly isothermal (constant sound speed $\cs$)
while Run~B is nearly isentropic ($\nabla-\nabla_{\rm ad}$ is
close to zero).
}\label{ppstrat}\end{figure}

We have evolved both simulations for several thermal and magnetic
diffusion times.
The mean stratification settles then to a new stratification (Fig.~\ref{ppstrat}).
Runs~A and B differ mainly in the mean temperature gradient which
becomes nearly constant in Run~A and stays finite for Run~B.
For Run~B we find that $\nabla-\nabla_{\rm ad}\approx-0.01$;
see Fig.~\ref{ppstrat}, so it is stably stratified, but close to marginal.
By contrast, in Run~B the stratification is strongly stable with
$\nabla-\nabla_{\rm ad}\approx-0.38$.

\begin{figure}[t!]\begin{center}
\includegraphics[width=\columnwidth]{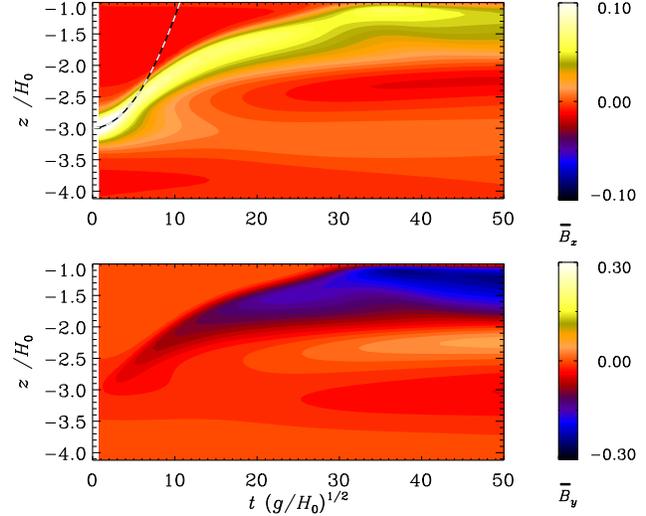}
\end{center}\caption[]{(online colour at: www.an-journal.org) 
Visualization of $\meanB_x$ and $\meanB_y$ as functions
of $t$ and $z$ for Run~A.
Note the initial ascent of $\meanB_x$ as well as the subsequent
amplification of $\meanB_y$.
For comparison, the curve $z=\half(\delta\rho/\rho)gt^2-z_0$
has been overplotted with an estimated value of $\delta\rho/\rho=0.2$.
}\label{pbxym_a_64o_TF}\end{figure}

When the simulation is started the initial flux tube begins to rise
owing to magnetic buoyancy, and it is also being sheared out in the
streamwise direction.
Both effects can clearly be seen in a $tz$ diagram of $\meanB_x$ and
$\meanB_y$; see \Fig{pbxym_a_64o_TF}.
Unless noted otherwise, an overbar denotes averaging over the
$x$ and $y$ directions.
The initial phase of the buoyant rise follows approximately a parabolic
trajectory, as is indicated by the superimposed dashed line in the first panel.

The effects of magnetic buoyancy can also be established at later times
through a systematic correlation between strong fields and upward motion.
We define a normalized buoyancy parameter,
\EQ
\mbox{Bu}=\bra{(\rho U_z)'\BB^2}/(\rho_0\Urms\Brms^2),
\EN
where the dash in $(\rho U_z)'$ denotes the departure of the mass flux
from the horizontal mean.
It turns out that $\mbox{Bu}$ is always positive.
This means that strong fields are systematically correlated with upward
motions and vice versa.
For Run~A this value is $\approx\!0.035$ and
for Run~B it is $\approx\!0.07$.
This supports the idea that magnetic buoyancy is indeed active,
but it is not clear that it still plays an important role at late
times compared to early times.
The peak values of Bu are around 3.5 at $t\approx5(g/H_0)^{1/2}$ both
for Runs~A and B.

\begin{figure*}[t!]\begin{center}
\includegraphics[width=\textwidth]{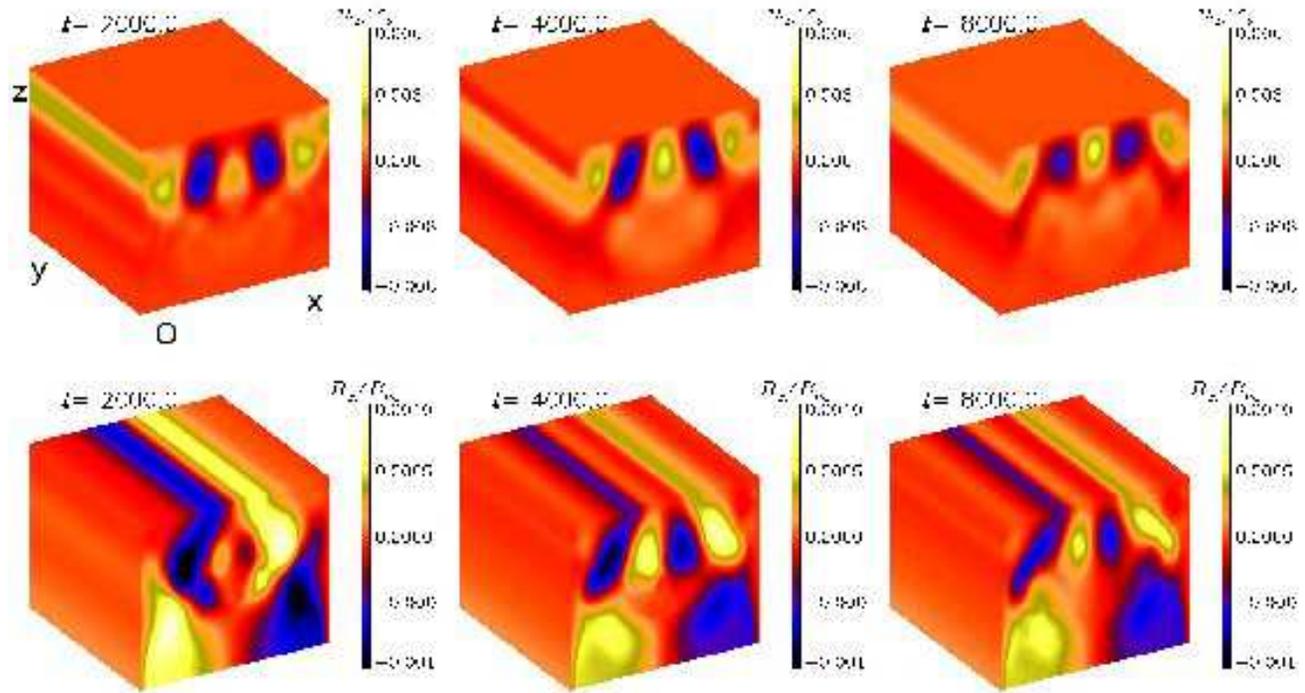}
\end{center}\caption[]{(online colour at: www.an-journal.org)
Visualization of $U_z$ (\emph{upper row}) and $B_z$ (\emph{lower row}) for Run~A at different times.
The coordinate directions are indicated in the upper left panel.
Note the symmetry of $U_z$ and the antisymmetry of $B_z$ with respect to $x=0$.
}\label{a_64a}\end{figure*}

\begin{figure*}[t!]\begin{center}
\includegraphics[width=\textwidth]{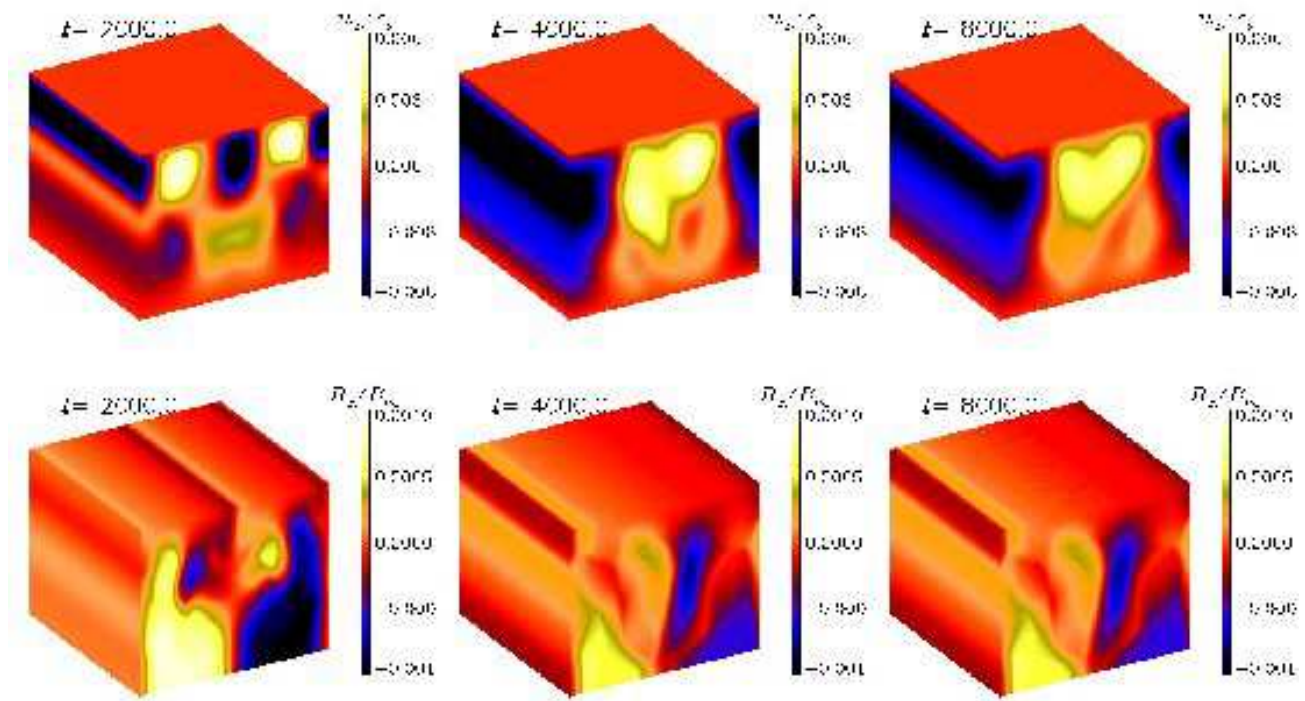}
\end{center}\caption[]{(online colour at: www.an-journal.org)
Same as \Fig{a_64a}, but for Run~B.
Here both $U_z$ and $B_z$ have no symmetry property with respect to $x=0$.
}\label{c1cT_64o}\end{figure*}

\begin{figure}[t!]\begin{center}
\includegraphics[width=\columnwidth]{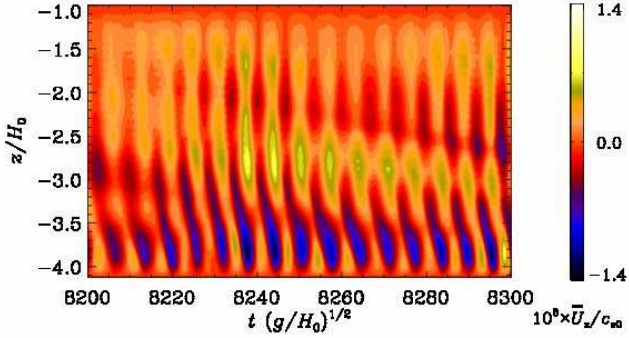}
\end{center}\caption[]{(online colour at: www.an-journal.org)
Horizontally averaged mean vertical velocity showing
persistent low amplitude oscillations for Run~A
at late times after $t=8200\,(H_0/g)^{1/2}$.
}\label{puzm}\end{figure}

In \Figs{a_64a}{c1cT_64o} we show the values of $U_z$ and $B_z$ on the
periphery of the computational domain for Runs~A and B at different times.
Toward the end of the simulation the state appears nearly perfectly steady,
but there are actually persistent low amplitude oscillations that remain
excited at all times.
In \Fig{puzm} we show a $zt$ diagram of the mean vertical velocity, $\meanU_z$.
The frequency of these oscillations is about $0.9\,(g/H_0)^{1/2}$,
This value lies between that of the Brunt-V\"ais\"al\"a frequency, whose
local value varies between 0.4 and 0.5 from bottom to top of the domain,
and the acoustic frequency of about $2\,(g/H_0)^{1/2}$.
Alfv\'en oscillations are also possible; their frequency is about 0.2
for Run~A in the final state.

\begin{figure}[t!]%\begin{center}
\includegraphics[width=\columnwidth]{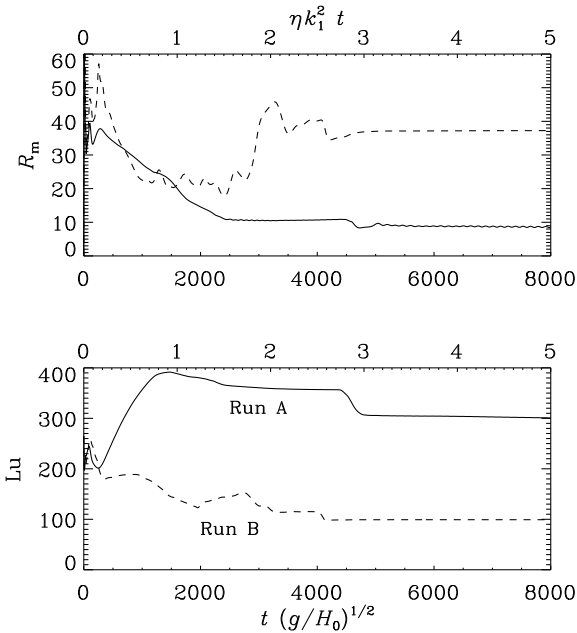}
%\end{center}
\caption[]{
Comparison of the evolution of rms velocity and magnetic field, normalized
in resistive units and expressed in terms of $\Rm$ and $\Lu$, for Runs~A and B.
The lower abscissa give time in dynamical units and the upper one in
resistive units.
}\label{pcomp_rm}\end{figure}

In \Fig{pcomp_rm} we plot the evolution of $\Urms$ and $\Brms$ in
diffusive time units, expressed in terms of the magnetic Reynolds
number ${\Rm=\Urms/\eta k_1}$ and Lundquist number ${\Lu=\vArms/\eta k_1}$,
respectively.
Here, $\vArms\!=\Brms/\sqrt{\mu_0\rho0}$ is the Alfv\'en speed.
Note that toward the end of the simulation we have $\Rm\approx10$
and $\Lu\approx300$.
There is no evident tendency for decay.

\begin{figure}[t!]\begin{center}
\includegraphics[width=\columnwidth]{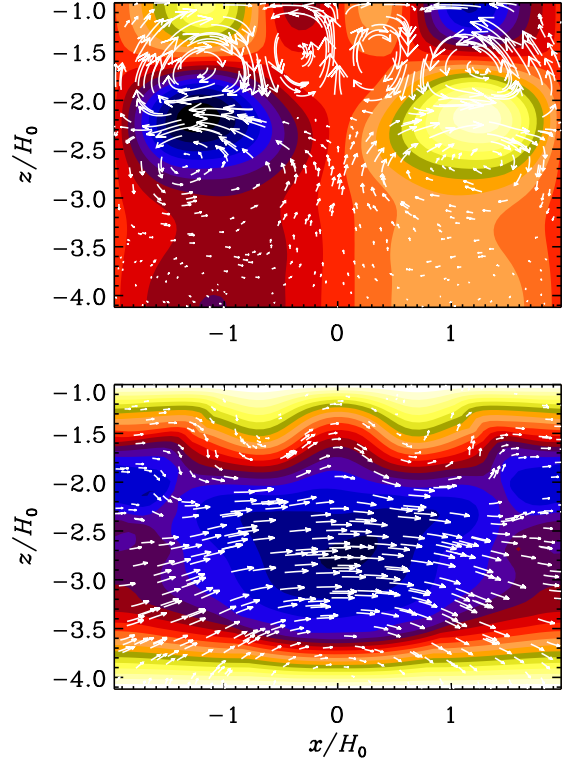}
\end{center}\caption[]{(online colour at: www.an-journal.org)
Zonally averaged velocity (upper panel) and magnetic field vectors
(lower panel) in the $xz$ plane superimposed on the corresponding
$y$ components of these fields for Run~A in the final and
nearly steady state.
Note the presence of a mean field with $\meanB_x>0$ and $\meanB_y<0$,
so $\meanB_x\meanB_y<0$, as expected for negative shear.
}\label{ppubxz_a_64a}\end{figure}

\begin{figure}[t!]\begin{center}
\includegraphics[width=\columnwidth]{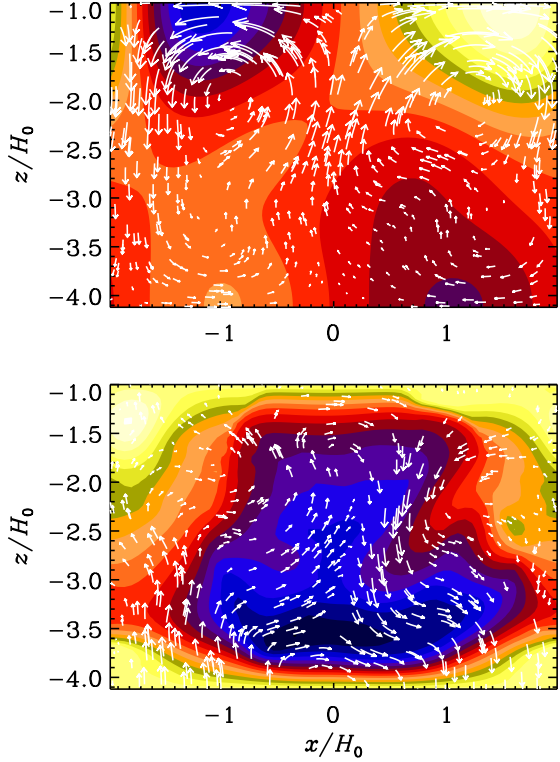}
\end{center}\caption[]{(online colour at: www.an-journal.org)
Same as \Fig{ppubxz_a_64a}, but for Run~B.
}\label{ppubxz_c1cT_64o}\end{figure}

%AB: \newpage commented out
%\newpage
\section{Flow of energy}

In order to understand where the energy comes from we consider here
the energetics of the system.
In a stratified hydromagnetic system there are four forms of energy:
potential, kinetic, thermal, and magnetic energy, defined respectively by
\EQ
E_{\rm P}=\int\rho\phi\dV,\quad
E_{\rm K}=\int \half\rho\UU^2 \dV,
\EN
\EQ
E_{\rm T}=\int \rho e \dV,\quad
E_{\rm M}=\int {\textstyle{1\over2\mu_0}}\BB^2\dV.
\EN
In the presence of stress-free boundary conditions,
the evolution of these energies is governed by the
following four ordinary differential equations:
\EQ
\dot{\rm E}_{\rm P}+W_{\rm b}=0,
\EN
\EQ
\dot{\rm E}_{\rm K}-W_{\rm b}-W_{\rm c}-W_{\rm L}-\epsilon^{\rm S}_{\rm K}+\epsilon_{\rm K}=0,
\EN
\EQ
\dot{E}_{\rm T}+W_{\rm c}-\epsilon_{\rm K}-\epsilon_{\rm M}-L_{\rm bot}+L_{\rm top}=0,
\EN
\EQ
\dot{E}_{\rm M}+W_{\rm L}-\epsilon^{\rm S}_{\rm M}+\epsilon_{\rm M}=0,
\EN
where dots denote time derivatives,
\EQ
W_{\rm b}=\int\rho\UU\cdot\grav\dV,\quad
W_{\rm c}=\int p\nab\cdot\UU\dV,
\EN
describe the work done by buoyancy and compression,
\EQ
W_{\rm L}=\int \UU\cdot(\JJ\times\BB)\dV,
\EN
is the work done by the Lorentz force,
\EQ
\epsilon^{\rm S}_{\rm K}=-\int\rho\UU(\SSSS\UU)\dV,\quad
\epsilon^{\rm S}_{\rm M}=\int\mu_0^{-1}\BB(\SSSS\BB)\dV,\quad
\EN
couple shear to kinetic and magnetic energy reservoirs,
\EQ
\epsilon_{\rm K}=\int2\rho\nu\SSSS^2\dV,\quad
\epsilon_{\rm M}=\int\sigma^{-1}\JJ^2\dV,
\EN
are viscous and Joule dissipation, and
\EQ
L_{\rm top}=\int_{\rm top}\FF_{\rm R}\cdot\dS,\quad
L_{\rm bot}=\int_{\rm bot}\FF_{\rm R}\cdot\dS,
\EN
are the luminosities at top and bottom of the domain.
The evolution of total energy, $E_{\rm tot}=E_{\rm P}+E_{\rm K}+E_{\rm T}+E_{\rm M}$,
is therefore governed by
\EQ
\dot{E}_{\rm tot}=\epsilon^{\rm S}_{\rm K}+\epsilon^{\rm S}_{\rm M}+L_{\rm bot}-L_{\rm top},
\EN
i.e.\ energy is supplied by shear coupling to kinetic and magnetic
energies, as well as by heating from below, and energy is removed by
radiation at the top.

\begin{figure}[t!]\begin{center}
\includegraphics[width=\columnwidth]{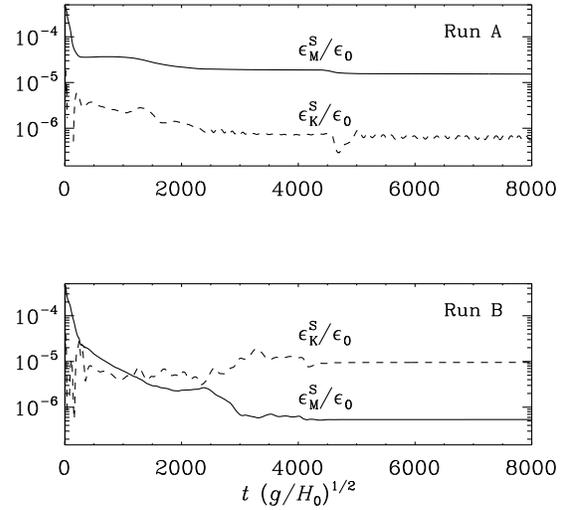}
\end{center}\caption[]{
Plot of the energy input from shear via the terms
$\epsilon^{\rm S}_{\rm K}$ and $\epsilon^{\rm S}_{\rm M}$.
Note the larger excess of $\epsilon^{\rm S}_{\rm M}$ compared with $\epsilon^{\rm S}_{\rm K}$
for Run~A compared with Run~B.
}\label{pcomp_gain}\end{figure}

\begin{figure}[t!]\begin{center}
\vskip-5mm
\includegraphics[width=\columnwidth]{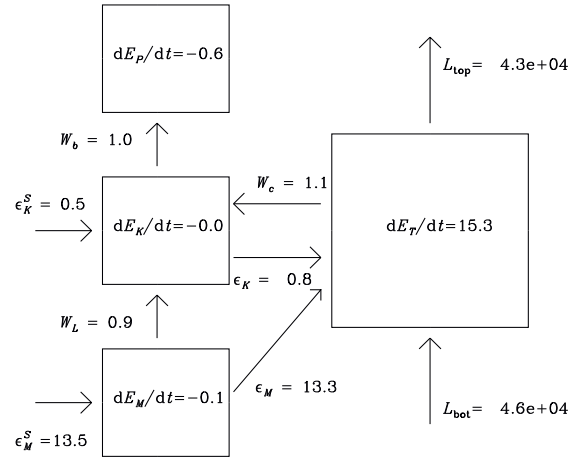}
\end{center}
\vskip-9mm
\caption[]{
Flow diagram showing the conversion between different energy forms
for Run~A.
All energy fluxes are normalized with respect to the averaged magnetic
energy content times the modulus of the shear rate and multiplied by
a factor of 1000 to bring the values closer to unity.
}\label{energybox_RunA}\end{figure}

\begin{figure}[t!]\begin{center}
\vskip-5mm
\includegraphics[width=\columnwidth]{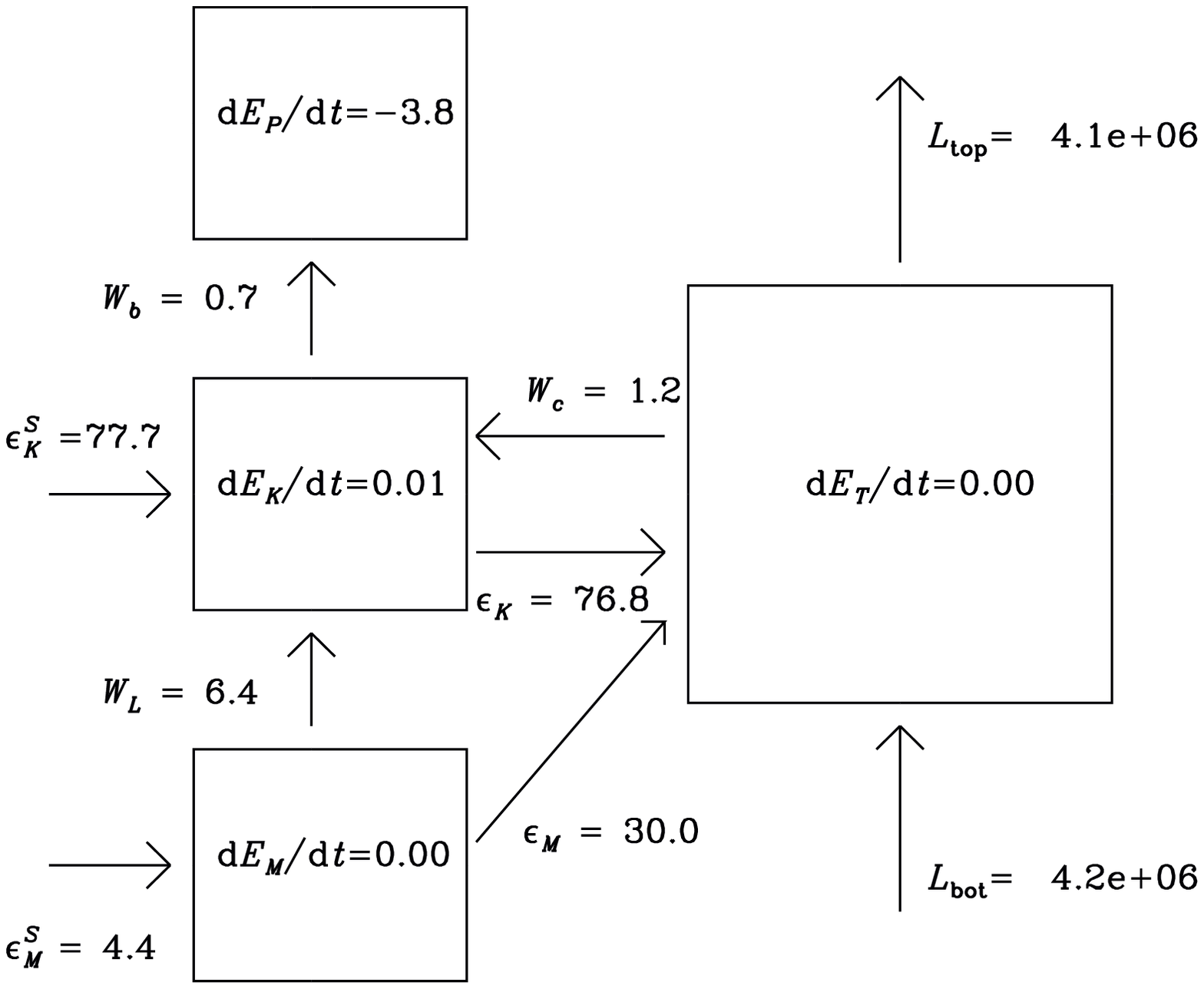}
\end{center}
\vskip-9mm
\caption[]{
Same as \Fig{energybox_RunA}, but for Run~B.
}\label{energybox_RunB}\end{figure}

Energy is exchanged and transformed into other types of energy,
as is shown in \Figs{energybox_RunA}{energybox_RunB} for Runs~A
and B, respectively.
It should be noted that, even if the system is not perfectly in
a steady state, we cannot expect perfect energy balance, because
the code solves the equations in non-conservative form.
However, the discrepancies should become smaller at larger
numerical resolution.
We notice that in all cases the energy flux at the top of the
domain is less than at the bottom, and that these values are much
larger than any of the other energy fluxes in the system.
This is related to the discretization error of the scheme and
does not indicate that energy is taken from the thermal energy flux.
For Run~B there is another obvious problem in that the magnetic energy
dissipation is much larger than what can be accounted for by the
influx of energy by shear.
This is related to the fact that in this run there is a fairly strong
magnetic field whose maintenance may be explained purely as the result
of numerical errors, even though no obvious problems can be seen from
images such as \Figs{a_64a}{c1cT_64o}.
This could be related to the fact that there are low amplitude
azimuthal variations with a relative amplitude of about $10^{-5}$
that cannot be seen unless the mean field is subtracted out.
In any case, we should keep in mind that the reason for the maintenance
of magnetic energy could be related to a very weak ``non-axisymmetry''
of the field that is related to numerical noise.
Regardless of these subtle shortcomings, there are many aspects that do
make sense.
In particular in Run~A the balance between inflow and outflow of magnetic
energy is quite reasonable, and so is the balance of kinetic energy
in Run~B.
This suggests that the flows in Run~A are mainly driven by work done by
the magnetic field against the shear, while in Run~B the driving comes
mostly from work done by the Reynolds stress against the shear.

\section{Profiles of mean field and velocity}

\begin{figure}[t!]\begin{center}
\includegraphics[width=\columnwidth]{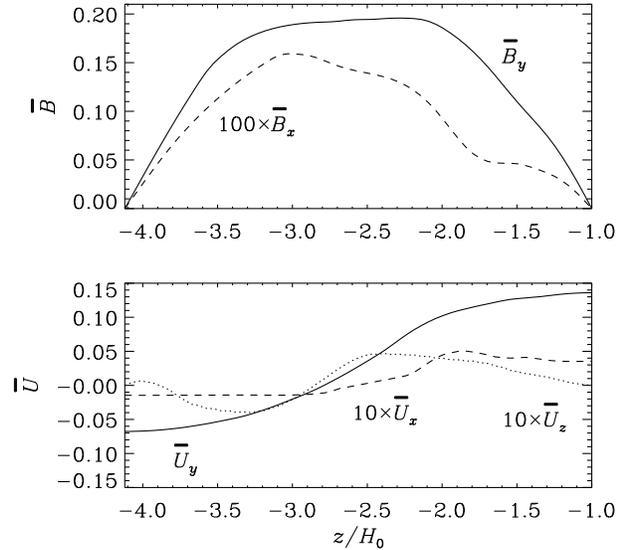}
\end{center}\caption[]{
$\meanB_y$ (solid line) together with $100\times\meanB_x$ (\emph{upper panel})
and $\meanU_y$ (solid line) together with $10\times\meanU_x$ and
$10\times\meanU_z$ (\emph{lower panel}), for Run~A, time averaged over
the last 1000 time units.
}\label{palpeta_ztm_a_64a_TF_BU}\end{figure}

It is interesting to confront the mean profile of the magnetic
field with stability criteria for magnetic buoyancy instabilities.
Following Newcomb (1961), a necessary condition for instability is
\EQ
\dd\ln|\BB|/\dd z > \gamma H_p N^2/v_A^2
\EN
(see also Hughes \& Proctor 1988).
This condition is actually met in the upper quarter of the simulation
domain where there is a rapid decrease of $|\BB|$ with height; see the
top panel of \Fig{palpeta_ztm_a_64a_TF_BU} for Run~A.

The mean ``toroidal'' velocity, $\meanU_y$ shows a systematic variation
in the $z$ direction that is reminiscent of that found for isotropically
forced shear-flow turbulence (K\"apyl\"a et al.\ 2009), which was then
interpreted in terms of a vorticity dynamo (Elperin et al.\ 2003).
In the present case, however, the magnetic field is already so strong
that such an effect would be suppressed.
Therefore, the flow is here more like a direct response to the magnetic
field.
The other two components of the mean flow are negligible by comparison.
However, we recall that the $z$ component of the mean velocity shows
persistent low amplitude oscillations (\Fig{puzm}) that are not present
in the other two components.

\section{Mean-field transport coefficients}

\begin{figure}[t!]\begin{center}
\includegraphics[width=\columnwidth]{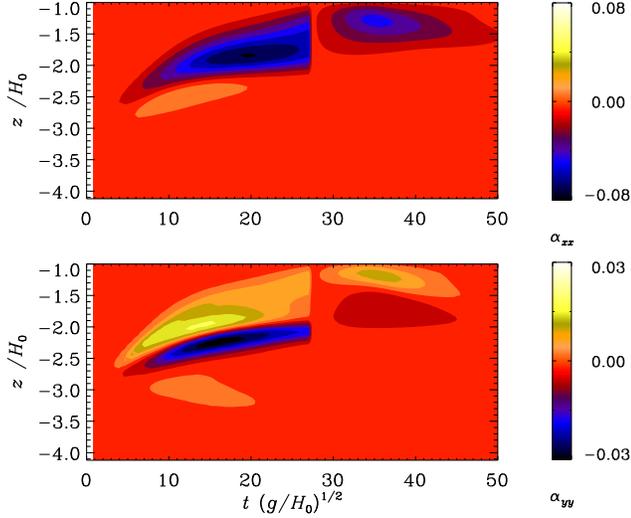}
\end{center}\caption[]{(online colour at: www.an-journal.org)
Diagonal component of the $\alpha$ tensor for Run~A at early times
(\emph{upper panel} is for $\alpha_{xx}$ and \emph{lower panel} for $\alpha_{yy}$).
Note that $\alpha_{yy}$ is negative in the lower part of the
initial rising flux tube and positive above, while  $\alpha_{xx}$
is negative throughout.
}\label{palpeta_zt_a_64o_TF}\end{figure}

In order to characterize the flow properties further we now consider
the ability of the flow to mix and to produce large-scale magnetic fields.
We do this by using the test-field method of Schrinner et al.\ (2005, 2007)
with sinusoidal test fields, as explained in detail in Brandenburg (2005)
and Brandenburg et al.\ (2008a,b).
In \Fig{palpeta_zt_a_64o_TF} we plot $\alpha_{xx}$ and $\alpha_{yy}$
versus time and height for Run~A at early times.
It turns out that $\alpha_{xx}$ is mainly negative at the location
of the initial rising flux tube while $\alpha_{yy}$ is negative in the
lower part of the tube and positive above.
In order to interpret this result we compare now with expectations
from mean-field theory.

\begin{figure}[t!]\begin{center}
\includegraphics[width=\columnwidth]{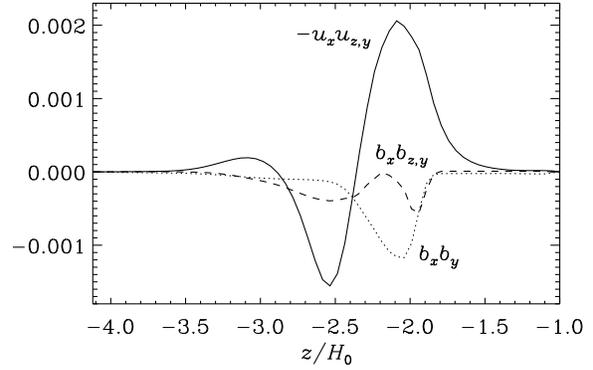}
\end{center}\caption[]{
Vertical profiles of $-\overline{u_{x}u_{z,y}}/g$ (solid line),
$\overline{b_{x}b_{z,y}}/g(\mu_0\rho_0)^{1/2}$ (dashed line), and
$\epsilon_{\rm buoy}\times\overline{b_{x}b_{y}}/\cs^2(\mu_0\rho_0)^{1/2}$
with $\epsilon_{\rm buoy}=0.1$ (dotted line) for Run~A at $t=10(H_0/g)^{1/2}$.
}\label{ppalpprof}\end{figure}

The combined presence of shear and magnetic buoyancy is particularly
interesting, because one may expect there to be an additional contribution
that is related to the magnetic stress, $\overline{b_xb_y}$ (Brandenburg 1998).
Such behavior was also reproduced using the second order correlation
approximation (R\"udiger \& Pipin 2000).
In the spirit of the $\tau$ approximation (e.g.\ Blackman \& Field 2003)
this term can be derived by calculating $\partial\meanEMF/\partial t$,
which has contributions from $\overline{\uu\times\dot{\bb}}$ and
$\overline{\dot{\uu}\times\bb}$.
The first term, which gives rise to the usual kinetic $\alpha$ effect, is
\EQ
(\overline{\uu\times\dot{\bb}})_y=\meanB_y\overline{u_{z,y}u_x}+...,
\EN
while the second term, which gives rise to the magnetic $\alpha$ effect
as well as a new term proportional to $\overline{b_xb_y}$.
To derive the second term we consider here the momentum equation in the form
\EQ
\dot{\uu}=...+\meanBB\cdot\bb/\rho\mu_0+{\delta\rho\over\rho}\grav,
\EN
where dots refer to additional terms that are less relevant for the
present discussion.
Replacing $\delta\rho/\rho$ by $\BB^2/2p\mu_0$ and linearizing about
$\meanB_y$, i.e.\ $\delta\rho/\rho\approx-\meanB_y b_y/p\mu_0$, yields
\EQ
(\overline{\dot{\uu}\times\bb})_y
=\meanB_y\left(\overline{b_{z,y}b_x}/\rho\mu_0
+g\overline{b_yb_x}/\mu_0 p\right)+...\, .
\EN
Therefore the $yy$ component of $\alpha$ should be
\EQ
\alpha_{yy}=-\tau\overline{u_xu_{z,y}}+
{\tau\over\mu_0\rho}\left(\overline{b_xb_{z,y}}
+\epsilon_{\rm buoy}\overline{b_xb_y}/H_0\right)+...,
\EN
where $\epsilon_{\rm buoy}$ is an empirical non-dimensional parameter
quantifying the relative importance of magnetic buoyancy effects.
In \Fig{ppalpprof} we plot the contributions from all three terms
for Run~A at $t=10(H_0/g)^{1/2}$.
It turns out that the main part of the vertical variation of $\alpha_{yy}$
is not determined by magnetic and buoyancy effects, but rather by the
kinematic contribution proportional to $-\overline{u_{x}u_{z,y}}$.

\begin{figure}[t!]\begin{center}
\includegraphics[width=\columnwidth]{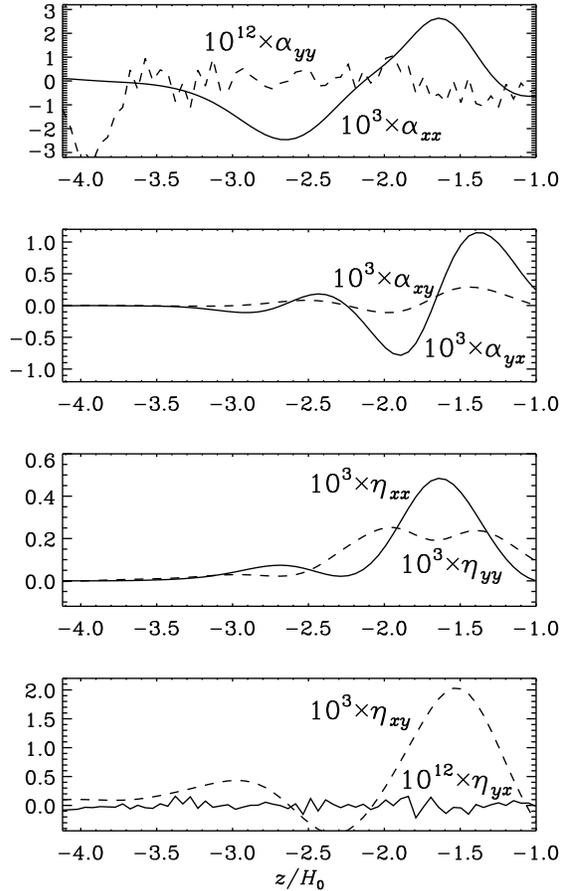}
\end{center}\caption[]{
All components of the $\alpha$ and $\eta$ tensors near the end of the run
for Run~A.
Note that $\alpha_{yy}$ and $\eta_{yx}$ are at the noise level and have
been multiplied by large factors ($\times10^{12}$) to see at least the noise.
}\label{palpeta_ztm_a_64a_TF}\end{figure}

The maintenance of the magnetic field over resistive time scales
can only be explained if there is a mean electromotive in the
$y$ direction that can balance the resistive losses of $\meanB_x$.
Those losses are proportional to $\eta\partial\meanB_x/\partial z$,
which gives the main contribution to $\meanJ_y$,
that is proportional to $\meanemf_y(z)$.
This is clearly suspicious and suggests that the fluctuations in the
$y$ direction are insufficient to explain the observed mean magnetic field.
This result is also in agreement with the fact that $\alpha_{yy}$
and $\eta_{xy}$, are found to be essentially around zero
(\Fig{palpeta_ztm_a_64a_TF}).
Thus, with the present knowledge we cannot propose a physical mechanism
for the maintenance of the observed mean field.

\begin{figure*}[t!]\begin{center}
\includegraphics[width=\textwidth]{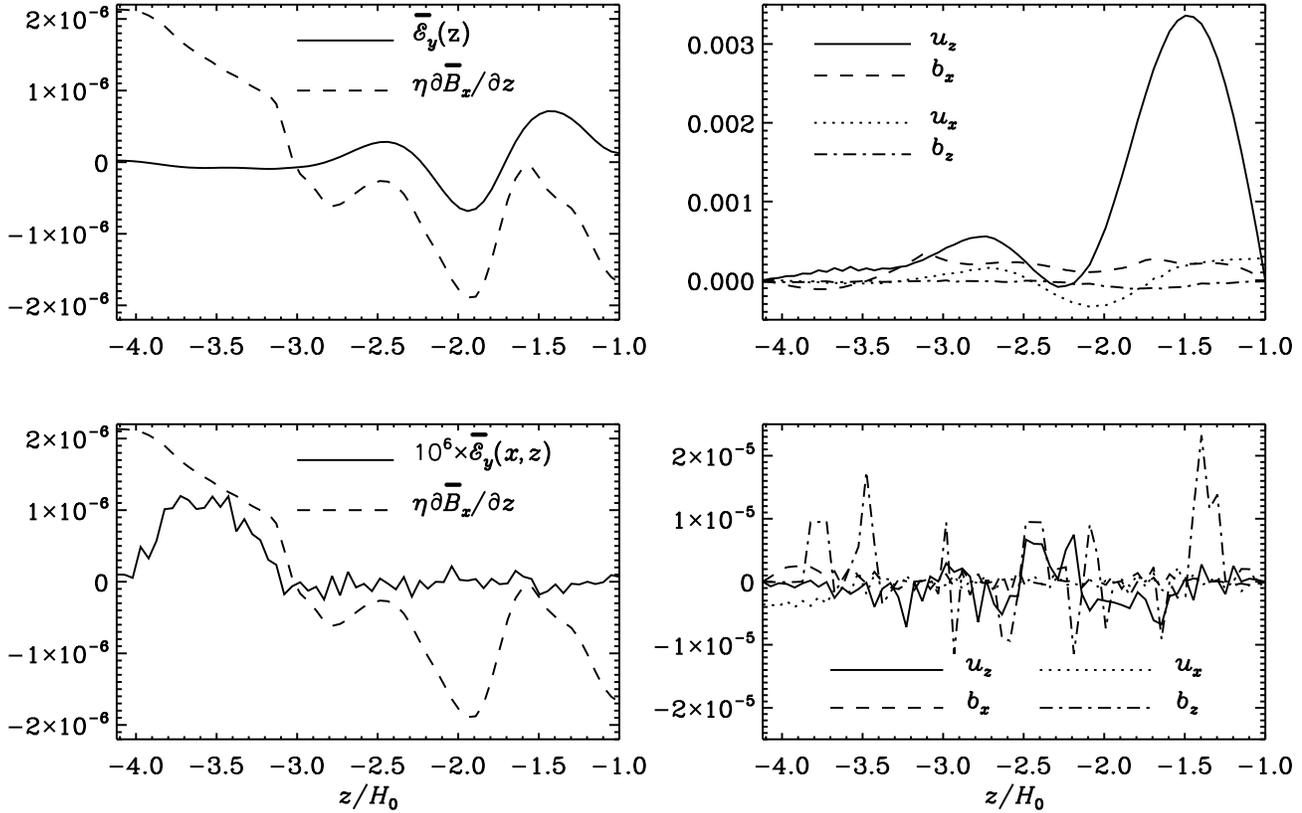}
\end{center}\caption[]{
Mean electromotive force in the $y$ direction overplotted with the
corresponding resistive losses (\emph{left panels})
together with plots showing the $z$ dependence of $u_z$ and $b_x$
as well as $u_x$ and $b_z$ for Run A.
For the upper row $xy$ averages were used while
for the lower row just $y$ averages have been employed.
}\label{pfluc_savefile2}\end{figure*}

In \Fig{pfluc_savefile2} (upper row) we plot $\meanemf_y(z)$ and confirm that
it is large enough to balance $\eta\partial\meanB_x/\partial z$,
which is overplotted by a dashed line.
The two curves are not in perfect agreement, but this could partly be explained
by the fact that the mean field is time-dependent and shows persistent
low-amplitude oscillations.
However, there is a caveat in that the definition of the fluctuating
quantities, $\uu$ and $\bb$ that enter $\meanEMF=\overline{\uu\times\bb}$,
have been computed using $xy$ averages.
If we use only $y$ averages, so that the mean field depends on both
$x$ and $y$, we find values that are essentially compatible with zero
(see the lower row of \Fig{pfluc_savefile2}).
Also the strength of fluctuations is rather low in that case.

\section{Conclusions}

In an attempt to study the possibility of dynamo action mediated by
magnetic buoyancy and driven by work done against an externally
imposed linear shear we came across a number of unusual phenomena.
Most surprising is probably the build-up and apparent maintenance of a
strong magnetic field.
The ratio of the Alfv\'en speed to the sound speed is about 0.1, while
the ratio of the Alfv\'en speed to the rms velocity of the motions is
about 10.
The magnetic field is nearly axisymmetric, i.e.\ independent of $y$.
In view of Cowling's theorem, this clearly raises doubts about the
validity of this result.
The possibility of numerical problems is supported by the fact that
at least for Run~B there is an unexplained imbalance between magnetic
energy input and output.
On the other hand, this rather obvious imbalance
applies only to Run~B, which is nearly isentropic.
For Run~A the overall energy balance is better.
Also, there are no obvious deficiencies that one would normally expect in the
case of insufficient numerical resolution such as oscillations on the
grid scale.

An important clue might be that the magnitude of the mean magnetic field
in the cross-stream direction can only be explained if the
$\overline{\uu\times\bb}$ correlation of the fluctuations are
evaluated as the departures from a horizontal $xy$ average.
If one takes just toroidal or $y$ averages, the resulting correlation
drops by six orders of magnitude into the numerical noise.
In this context it should be noted that our calculations are normally
performed in single precision, which has been sufficient for most previous
applications.
The present simulations may present an exception to this.

Yet another possible clue comes from the fact that our models
exhibit sustained oscillations.
They are most prominent in the isothermal case (Run~A) and less
prominent in the nearly isentropic case (Run~B).
This may suggest that buoyancy oscillations could contribute to
driving the mean electromotive force responsible for sustaining
the mean field $\meanB_x$ in the cross-stream direction.
(The mean field in the streamwise direction, $\meanB_y$, is
readily explained by the shear.)
On the other hand, an analysis of the flow in terms of the test-field
method reveals that both $\alpha_{yy}$ and $\eta_{yx}$ are essentially
zero in the final state, making it impossible to explain the magnetic
field as a result of an $\alpha$ effect or a shear--current effect.

In any case, it is clear that shear-driven flows in
stratified systems can exhibit rich behavior.
Magnetic buoyancy effects are clearly seen at early times shortly
after injecting the initial cross-stream magnetic field.
Within the present setup we have not seen the vigorous dynamo action
with visibly nonaxisymmetric fields reported by Cline et al.\ (2003).
This could simply be related to differences in the parameters or to
differences in the flow geometry.
We recall that in Cline et al.\ (2003) there was a non-shearing upper
part that is absent in the present work.
In addition, they used a periodic shear profile rather than a linear one.
In any case, it will be worthwhile performing new simulations using perhaps
also other methods and certainly larger resolution.
It may be worthwhile to continue these studies using
an isothermal setup rather than a nearly isentropic one,
provided the oscillations seen in the present work are indeed
an important element of the overall dynamics.

%AB: \newpage commented out
%\newpage
\acknowledgements
We thank the referee for pointing out several shortcomings in the
original version of the paper.
We acknowledge the use of computing time at the Center for
Parallel Computers at the Royal Institute of Technology in Sweden.
This work was supported in part by
the European Research Council under the AstroDyn Research Project 227952
and the Swedish Research Council grant 621-2007-4064.

\end{document}